\documentclass[11pt]{article}
\def\be{\begin{equation}}
\def\ee{\end{equation}}
\def\ba{\begin{eqnarray}}
\def\ea{\end{eqnarray}}
\def\la{\langle}
\def\ra{\rangle}
\def\a{\alpha}

\def\h{\hskip 1cm}

\def\lo{\longrightarrow}
\def\A1{A_{-1}}
\usepackage{epsfig}
\begin{document}
\begin{titlepage}
\vspace{4cm}
\begin{center}{\Large \bf Entanglement and quantum phase transitions in matrix product spin one chains}\\
\vspace{2cm}\h S.
Alipour$^{\sharp}$\footnote{email:salipour@physics.iust.ac.ir},\h
V. Karimipour$^\dag$ \footnote{email:vahid@sharif.edu},\h L.
Memarzadeh$^\ddag$\footnote{Corresponding author, email:laleh@physics.sharif.edu}, \\
\vspace{1cm} $^{\dag,\ddag}$Department of Physics, Sharif
University of Technology, \\P.O. Box 11365-9161, Tehran, Iran

\vspace{1cm} $^{\sharp}$Department of Physics, Iran University of
Science and Technology,\\ Narmak, P.O. Box 16765-163, Tehran, Iran\\
\end{center}
\vskip 3cm
\begin{abstract}
We consider a one-parameter family of matrix product states of
spin one particles on a periodic chain and study in detail the
entanglement properties of such a state. In particular we
calculate exactly the entanglement of one site with the rest of
the chain, and the entanglement of two distant sites with each
other and show that the derivative of both these properties
diverge when the parameter $g$ of the states passes through a
critical point. Such a point can be called a point of quantum
phase transition, since at this point, the character of the
matrix product state which is the ground state of a Hamiltonian,
changes discontinuously. We also study the finite size effects
and show how the entanglement depends on the size of the chain.
This later part is relevant to the field of quantum computation
where the problem of initial state preparation in finite arrays
of qubits or qutrits is important. It is also shown that
entanglement of two sites have scailing behavior near the
critical point.
\end{abstract}
\end{titlepage}

\section{Introduction}
Interacting spin systems are among the most promising candidates
for the actual implementation of quantum computers in the future.
In the new terminology which has emerged since the upsurge of
interest in quantum computation, a spin 1/2 system refers not
only to the actual spin degrees of freedom of an atom or nucleus,
but it refers to any system, with any number of levels, in which
we have selected two states for encoding the information. For
example the ground state and the first excited state of an ion in
an ion trap, make a spin 1/2 system or a so called qubit. Similar
terminology is used for three level states or qutrits. These
latter systems and their generalizations to $d-$ level states or
qudits are of immense interest, since it is not yet clear if
actual implementation of quantum computers will be based on two
level systems. In view of this, many systems which have been
traditionally the focus of study in condensed matter physics are
being examined from different points of view related to quantum
computation and information. The most important property of an
interacting spin system, which is of relevance to this new
emerging field, is entanglement or non-local quantum correlation.
In fact it is considered as a resource, like energy, since it
plays a vital role in any process of quantum information and
computation, moreover it can be measured, manipulated and
transferred. Consider the ground state of an interacting spin
system, comprising $N$ spins. There are some basic questions
regarding entanglement: How much two distant spins are entangled
with each other? How this entanglement varies with the system
size $N$? Is there any threshold distance, beyond which there
will be no entanglement at all? How the entanglement varies when
we approach a point of quantum phase transition? Answering these
questions requires tools which have been developed only recently
in the field of quantum information \cite{nielsen, osterloh,
occonersWooters, bosevedral,ak,Zanardi}. In this way a fruitful
field of investigation at the borderline of condensed matter
physics and quantum information has emerged. The aim of this
letter is to investigate such questions for a class of spin-one
states, which are known to be exact ground states of certain
multi-parameter families of Hamiltonians describing nearest
neighbor interactions of spin-one particles on a periodic chain.
The method for construction of such states, known also as the
matrix product formalism, was first introduced in
\cite{Affleck,Fannes,Z1} and then applied to various models in
\cite{Z2,Z3,RO1,RO2}. Recently it has been applied even to two
dimensional models \cite{Z4,AKS}. Such states can be constructed
to have specified symmetry properties or even to induce certain
kinds of quantum phase transitions with pre-determined
properties\cite{Cirac}. The entanglement properties of the so
called AKLT models \cite{Affleck}, which inspires the matrix
product states was first studied in \cite{Korepin}. Here we study
the entanglement properties of a one parameter deformation of AKLT
models.\\
We first study the entanglement of one site with the rest of the
chain and the entanglement of two distant sites in the
thermodynamic limits, however we mainly focus on states with
finite but arbitrary number of spins, since this is the case of
interest from the point of view of quantum computation and
information. We will determine the entanglement of two spins in
the lattice as a function of their distance, the coupling
constant of the state or the Hamiltonian, which we denote by $g$,
and the system size. The results are that: 1- When the parameter
$g$ approaches its critical value $g_0=0$, the range of
entanglement increases indefinitely at the cost of its magnitude,
fig. (\ref{Neg_g_r}), 2- for any non-zero value of $g$, there is
a threshold distance beyond which there is no entanglement
between spins, fig. (\ref{Neg_r}), 3- for any two spins with a
fixed distance, there is a threshold system size, above which
entanglement vanishes, fig. (\ref{Neg_N}), and there is a scaling
behaviour in the entanglement of two adjutant spins which is
shown in figure (\ref{All2}).\\
Let us first remind the matrix product formalism in a language
which we find convenient.

\section{Matrix Product States}\label{MPS}
For a homogeneous ring of $N$ sites, where each site describes a
$d-$level state. The Hilbert space of each site is spanned by the
basis vectors $|i\ra, \ \ i=0,\cdots d-1$. A state
\begin{equation}\label{state}
    |\Psi\ra=\sum_{i_1,i_2,\cdots i_N}\psi_{i_1i_2\cdots
    i_N}|i_1,i_2,\cdots, i_N\ra
\end{equation}
is called a matrix product state, if there exist matrices $A_i,\ \
i=0\cdots d-1$ (of dimension $D$) such that
\begin{equation}\label{mat}
    \psi_{i_1i_2\cdots
    i_N}=\frac{1}{\sqrt{Z}}tr(A_{i_1}A_{i_2}\cdots A_{i_N}),
\end{equation}
where $Z$ is a normalization constant equal to $
    Z=tr(E^N)
$ and $ E:=\sum_{i=0}^{d-1} A_i^*\otimes A_i. $

The correlation functions are readily calculated in this
formalism. For example, for the one-point functions we have
\begin{equation}\label{1point}
    \la \Psi|O_k|\Psi\ra = \frac{tr(E^{k-1}E_O
    E^{N-k})}{tr(E^N)},
\end{equation}
where $E_O=\sum_{i,j}\la i|O|j \ra A_i^*\otimes A_j.$ In the
thermodynamic limit, only the largest eigenvalue of $E$ survives
and so any level crossing in the largest eigenvalue of $E$, leads
to a discontinuity of correlation functions. This may be termed an
MPS-quantum phase transition \cite{Cirac}.
\subsection{Gauge Transformations}
From (\ref{mat}), it is evident that two sets of matrices
$\{A_i\}$ and $\{A'_i\}$ lead to the same matrix product state if
they are related as $ \  A'_i = \mu S A_i S^{-1},\ $ where $\mu$
is a scale factor and $S$ is any invertible matrix. Actually the
gauge transformation can be more general than this, namely
$A'_i=\mu SA_iS'$ with $S'S=I$. Such transformations can be used
to gauge away irrelevant parameters in the matrices $A_i$.
\subsection{Symmetries}
On a ring, the state (\ref{state}) is invariant under translation.
Demanding more symmetries imposes constrains on the matrices
$A_i$. Considering equation (\ref{mat}), the state is symmetric
under parity if there exists a matrix $\Pi$ such that
\begin{equation}\label{parity}
A_i^T=\sigma\Pi A_i\Pi^{-1} \hskip 1cm \sigma=\pm1
\end{equation}
where $A^T$ is the transpose of $A$ and it has time reversal
symmetry if the matrices $A_i$ are real.\\
Consider now a local symmetry operator $R$ acting on a site as
$R|i\ra=R_{ji}|j\ra$ where summation convention is being used. $R$
is a $d$ dimensional unitary representation of the symmetry. A
global symmetry operator ${\cal R}:=R^{\otimes N}$ will then
change this state to another matrix product state
\begin{equation}\label{mpsPrime}
    \Psi_{i_1i_2\cdots i_N}\lo \Psi':=tr(A'_{i_1}A'_{i_2}\cdots
    A'_{i_N}),
\end{equation}
where $ A'_i:=R_{ij}A_j.$ The state $|\Psi\ra$ is invariant under
this symmetry if there exists an operator $U(R)$ such that
\begin{equation}\label{symm}
    R_{ij}A_j=U(R)A_iU^{-1}(R).
\end{equation}
Thus $R$ and $U(R)$ are two unitary representations of the
symmetry, respectively of dimensions $d$ and $D$. In case that
$R$ is a continuous symmetry with generators $T_a$, equation
(\ref{symm}), leads to
\begin{equation}\label{symmalg}
    (T_a)_{ij} A_j=[{\cal T}_a,A_i],
\end{equation}
where $T_a$ and ${\cal T}_a$ are the $d-$ and $D-$dimensional
representations of the Lie algebra of the symmetry.
\subsection{The Hamiltonian}
To construct a Hamiltonian  with nearest neighbor interaction, in
a way that the state in equation (\ref{state}) be its ground
state, we have to find the null space of the density matrix of
two adjacent sites which is given by:
\begin{equation}\label{rhok}
    \rho_{ij;kl}=\frac{tr((A_{i}^* A_{j}^*\otimes A_{k}A_{l})E^{N-2})}{tr(E^N)}.
\end{equation}
The null space of this reduced density matrix include the
solutions of
\begin{equation}\label{cc}
    \sum_{k,l}^{d-1}c_{kl}A_{k}A_{l}=0.
\end{equation}
The number of independent solutions of this system of equation is
$d^2-D^2$. Thus for this density matrix to have a null space it is
sufficient that $ d\
>\ D. $
Let the null space of the reduced density matrix be spanned by the
orthogonal vectors $|e_{\a}\ra, \ \ \ (\a=1, \cdots s\geq
d^2-D^2)$. Then we can construct the local hamiltonian acting on
$2$ consecutive sites as $ h:=\sum_{\a=1}^s \lambda_{\a}
|e_{\a}\ra\la e_{\a}|,$ where $\lambda_{\a}$'s are non-negative
constants. These parameters together with the parameters of the
vectors $|e_\a\ra $ inhertited from those of the original
matrices $A_i$, determine the total number of coupling constants
of the Hamiltonian. The full Hamiltonian on the chain is written
as $ H=\sum_{l=1}^N h_{l,l+1}. $, where $h_{i,i+1}$ is the
embedding of $h$ into sites $i$ and $i+1$. The state $|\Psi\ra$
is then a ground state of this Hamiltonian with vanishing energy
(for an exposition see \cite{AKS}).
\section{Matrix Product States on Spin 1 Chains}
\subsection{Construction of the state}
The matrix product state, that we use for spin-one systems, has
already been constructed in \cite{Z2}. Here we review its
construction in the language introduced in previous section for
completeness. Since
 $d=3$, to guarantee a null space for the two-site density matrix,
we set $D=2$. The matrices $A_1$, $A_0$ and
$A_{\overline{1}}\equiv A_{-1}$ correspond to the local states
$|1\ra$, $|0\ra$ and $|\overline{1}\ra\equiv|-1\ra$ respectively,
where $S_z|m\ra=m|m\ra$. Considering equation (\ref{symmalg}), the
symmetry around the $z$ axis requires that
\begin{equation}\label{Sz}
    [S_z,A_m]=mA_m,
\end{equation}
where $S_z=\frac{1}{2}\left(\begin{array}{cc} 1 & 0
\\ 0 & -1\end{array}\right)$. Solving (\ref{Sz}), demanding
parity symmetry (\ref{parity}) and getting rid of irrelevant
parameters with suitable gauge transformations, leave us with:
\begin{equation}\label{Ai}
{\cal A} =\left(\begin{array}{c c}
     |0\ra &-\sqrt{2}g|1\ra\\
     |\overline{1}\ra &\sigma |0\ra
     \end{array}\right),
\end{equation}
where we have used the compact notation ${\cal A}:=\sum_i A_i
|i\ra$ . Note that the state constructed in this way
automatically has spin-flip symmetry, i.e. $XA_mX^{-1}=\sigma
A_{\overline{m}}  $ with $ X=\left(\begin{array}{c c}
     0&-\sigma g\\
     1&0
     \end{array}\right)
$. Also at $(g,\sigma)=(1,-1)$, the so-called AKLT point
\cite{Affleck}, the model has full rotational symmetry, since in
this case the states $-\sqrt{2}\sigma_+, \sigma_z$ and
$\sqrt{2}\sigma_-$,
form a vector under the adjoint representations of the rotation group. \\
\subsection{The Hamiltonian}
As mentioned previously, we must solve (\ref{cc}) for matrices
(\ref{Ai}) to construct the hamiltonian. It is straightforward to
verify that the solution space of (\ref{cc}) is spanned by the
following vectors:
\begin{eqnarray}\label{}
 |e_1\ra&=& |1,1\ra\cr
 |e_2\ra&=& \frac{1}{\sqrt{2}}(|1,0\ra-\sigma|0,1\ra)\cr
 |e_3\ra&=&\frac{1}{\sqrt{2+4g^2}}(|1,\overline{1}\ra+2g|0,0\ra+|\overline{1},1\ra)\\
 |e_4\ra&=&\frac{1}{\sqrt{2}}(|0,\overline{1}\ra-\sigma|\overline{1},0\ra\ra)\cr
 \nonumber|e_5\ra&=& |\overline{1},\overline{1}\ra.
\end{eqnarray}
With these vectors, we write the local Hamiltonian as
\begin{equation}\label{h}
h=a(|e_1\ra\la e_1|+|e_5\ra\la e_5|)+b (|e_2\ra\la e_2| +
|e_4\ra\la
    e_4|)+c|e_3\ra\la e_3|,
\end{equation}
 to preserve the $Z_2$ symmetries mentioned above. Writing this
in terms of local spin operators, the final form of $H$ is
obtained as:
\begin{eqnarray}\label{H}
H=\sum_{i=1}^N J_1 {\bf S}_i\cdot {\bf S}_{i+1} + J_2 ({\bf
S}_i\cdot {\bf S}_{i+1})^2 + J_3 S_{z,i}S_{z,i+1}\cr + J_4
(S_{z,i}S_{z,i+1})^2 + J_5 S_{z,i}^2 + J_6 \{{\bf S}_i\cdot {\bf
S}_{i+1},S_{z,i}S_{z,i+1}\}_+,
\end{eqnarray}
where
\begin{eqnarray}\label{J}
    J_1&=&-b\sigma(1+2g^2),\ \ \ \ \  J_2= c,\cr
    J_3&=& (a+b\sigma)(1+2g^2),\\
    J_4&=& (a+2b(\sigma-1))(1+2g^2)+(1+2g)^2c,\cr
    J_5&=& 2 b(1+2g^2)+2c(1-4g^2),\cr
\nonumber J_6&=&  -b\sigma(1+2g^2)-c(1+2g).
\end{eqnarray}
In writing the above Hamiltonian we have ignored an overall
additive constant and have re-scaled the operator (\ref{h}) by a
constant $2(1+2g^2)$.  This represent a four parameter family of
Hamiltonians which have (\ref{state}) as their ground state. Full
rotational symmetry exists ,when we have $(g,\sigma)=(1,-1)$, and
$a=b=c=1$, for which case the Hamiltonian is known as the AKLT
model.

\section{The thermodynamic limit}
We can derive many properties of the ground state using the
transfer matrix formalism explained in section (\ref{MPS}).
 The eigenvalues of the matrix $E$ are found to be
\begin{equation}\label{eigenE}
\lambda_1=1+2g\h \lambda_2=1-2g\h  \lambda_{3,4}=\sigma.
\end{equation}
This shows a level crossing in the largest eigenvalue of $E$ and
hence a singularity in correlation functions at $g=0$. The average
magnetization and the correlation functions are found to be
\cite{Z2}
$$<S_x^i>=<S_y^i>=<S_z^i>=0,$$
and
$$<S_z^1S_z^{r}>=-\frac{4g^2}{(1-2|g|)^2}\left(\frac{1-2|g|}{1+2|g|}\right)^r$$
$$<S_{{\bf n}}^1S_{{\bf n}}^{r}>=-2|g|(\sigma-Sign(g))\left(\frac{\sigma}{1+2|g|}\right)^r,$$
where ${\bf n}$ is any unit vector in the $xy$ plane. The
longitudinal and transverse correlation length, diverges at $g=0$,
It is a natural question to ask if the same thing happens for
entanglement when $g$ approaches this critical point. In
\cite{indian}, the one-site entropy which measures the
entanglement of one site with the rest of the lattice and also
the two-site entropy which measures the entanglement of these two
sites with the rest of the lattice were calculated in the
thermodynamic limit. Here  we use the negativity to measure how
much two distant spins are entangled with each other. Moreover,
we study in detail finite size effects to see how various
properties of entanglement depend on the system size.

\subsection{Entanglement of two distant sites}
In the ground state of (\ref{H}), any two particles will be in a
mixed state. The reduced density matrix of two spins located at
sites $1$ and $r$ is denoted by $\rho^{1,r}$. The rotational
symmetry around the $z$ axis shows that $\rho_{ij;kl}=0$ unless
$i+j=k+l$. Also the parity symmetry entails the condition
$\rho_{ij,kl}=\rho_{\overline{i}\overline{j};\overline{k}\overline{l}}$,
where $\overline{i}=-i$. Straightforward calculation shows that
\begin{equation}\label{r2Thermo}
\rho(1,r)=\left(\begin{array}{c c c c c c c c c}
\alpha& & & & & & & & \\
 &|g|\gamma& &\mu& & & & & \\
 & &\beta& &\delta& &\nu & & \\
 &\mu& &|g|\gamma& & & & & \\
 & &\delta& &\gamma& &\delta& & \\
 & & & & &|g|\gamma& &\mu& \\
 & &\nu& &\delta& &\beta & & \\
 & & & & &\mu& &|g|\gamma& \\
 & & & & & & & &\alpha\\
\end{array}\right)
\end{equation}
in which
\begin{eqnarray}\label{elements}
\left(\begin{array}{c} \a
\\ \beta \end{array}\right)&=&\frac{g^2(\lambda_1^{r-2}\mp
\lambda_2^{r-2})}{\lambda_1^r},\hskip 0.7cm \nu = 0, \ \ \ \
\gamma=\frac{1}{\Lambda_1^2}, \cr \cr
\delta&=&-g\left(\frac{\sigma}{\Lambda_1}\right)^r, \hskip 1.2 cm
\mu=\sigma|g|\left(\frac{\sigma}{\Lambda_1}\right)^r,
\end{eqnarray}
where $\Lambda_1=1+2|g|$ and $\Lambda_2=1-2|g|$. Since the state
of two sites, is a mixed state, we can not use von Neumann
entropy to measure the entanglement between these two sites. In
\cite{Peres} it is shown that the necessary condition for a mixed
state $\rho$, to be separable is that its partial transpose has
non-negative eigenvalues. The quantitative version of this
criterion is \textit{Negativity} which is defined as follows
\cite{V-W}:
\begin{equation}
\mathcal{E}(\rho(1,r))=\frac{||\rho^{T_A}(1,r)||_1-1}{2}
\end{equation}
where $\rho^{T_A}(1,r)$ is the partial transpose of $\rho(1,r)$
with respect to the subsystem $A$, and $||X||_1=\sqrt{X^{\dag}X}$
is the trace norm of $X$. Equivalently it is equal to the sum of
absolute values of negative eigenvalues of the matrix
$\rho^{T_A}(1,r)$. In other words by means of Negativity we can
measure the degree to which the partial transpose of the state
$\rho$ fails to be positive or how far the state of the two
particles is from a separable state. A basic property of
Negativity is that it is an entanglement monotone, meaning that
the more entangled  a state, the more negative it is in the above sense.  \\
The eigenvalues of $\rho^{T_A}(1,r)$ are found from
(\ref{r2Thermo}) to be:
\begin{eqnarray*}
\omega_1&=&\alpha \h \omega_{2,3}=\beta \h
\omega_{4,5}=|g|\gamma+\delta\\ \omega_{6,7}&=&|g|\gamma-\delta,
\ \
\omega_{8,9}=\frac{\alpha+\gamma+\nu}{2}\pm\frac{1}{2}\sqrt{(\alpha-\gamma+\nu)^2+8\mu^2}.
\end{eqnarray*}

\begin{figure}
\centering
\includegraphics[width=8cm,height=8cm,angle=-90]{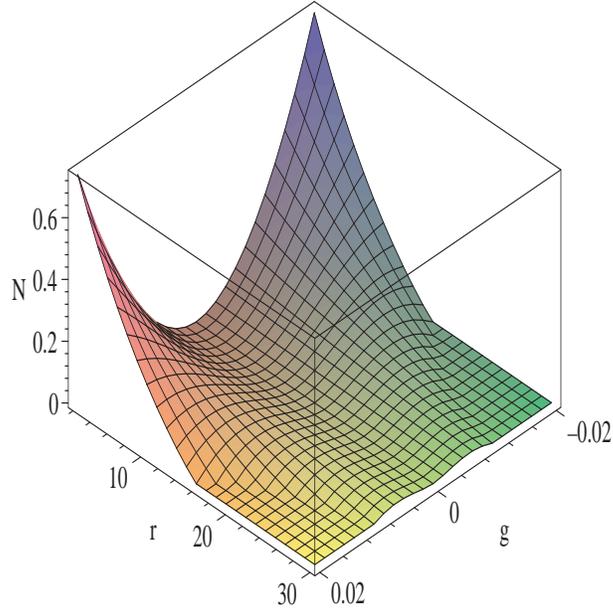}
    \caption{(Color Online) The entanglement of two spins at sites 1 and $r$ as a
    function of their distance and the coupling $g$. Near the critical
    point the range of entanglement increases at the cost of its amplitude,
    for every $g$ there is a maximum distant, beyond which there is no entanglement.}
    \label{Neg_g_r}
\end{figure}
From these eigenvalues the negativity and hence the entanglement
of two spins can be calculated. Figure (\ref{Neg_g_r}) shows the
entanglement of two spins, one at site $1$ and the other at site
$r$ as a function of $r$ and the parameter $g$. It is seen that
in these models the range of entanglement increases as we
approach the critical point, however its value decreases when we
approach this point. We can obtain an approximate expression for
the maximum range of entanglement. Inspection shows that the only
eigenvalue of $\rho^{T_A}(1,r)$ which may go negative is
$$\omega_9=\frac{\alpha+\gamma}{2}-\frac{1}{2}\sqrt{(\alpha-\gamma)^2+8\mu^2},$$
note that $\nu=0$ in the thermodynamic limit (\ref{elements}).
Thus the inequality $\omega_9<0$ determines the range of
entanglement. This is however equivalent to $ \a\gamma < 2\mu^2$
or in view of (\ref{elements})
$$(1+2|g|)^{r-2}-(1-2|g|)^{r-2}< 2(1+2|g|)^{2-r}.$$
For small values of $g$, i.e. $|g|\ll 1$, we can write $1\pm
2|g|\approx e^{\pm 2|g|}$ and then the above inequality transforms
to
$$r\leq \frac{ln(3)}{4 |g|}+2,$$
implying that a entanglement exists up to a range of $r_0\approx
\frac{ln (3)}{4 g}+2$. As an example for $g=0.02$, this gives
$r_0\approx 15$, which is also evident from figure
(\ref{Neg_g_r}).

\section{Systems with finite size}
The interest in entanglement properties of spin systems stems not
only from its possible relation to the critical properties of such
systems, but also from their possible candidacy for the future
implementation of quantum computers. In this case we are dealing
with a finite array of interacting spins which has relaxed to its
ground state. Tuning the interactions of these spins with each
other, changes the ground state and it is desirable to have
controllable entanglement between different spins of this array.
Therefore in this section we study more closely the properties of
such matrix product states for finite values of $N$.\\

%%%%%%%%%%%%%%%%%%%%%%%%%%%%%%%%%%%%%%%%%%%%%%%%%%%%%%%%%%%%%%%%%%%%%%%%%
\begin{figure}
\includegraphics[width=14cm,height=7cm,angle=0]{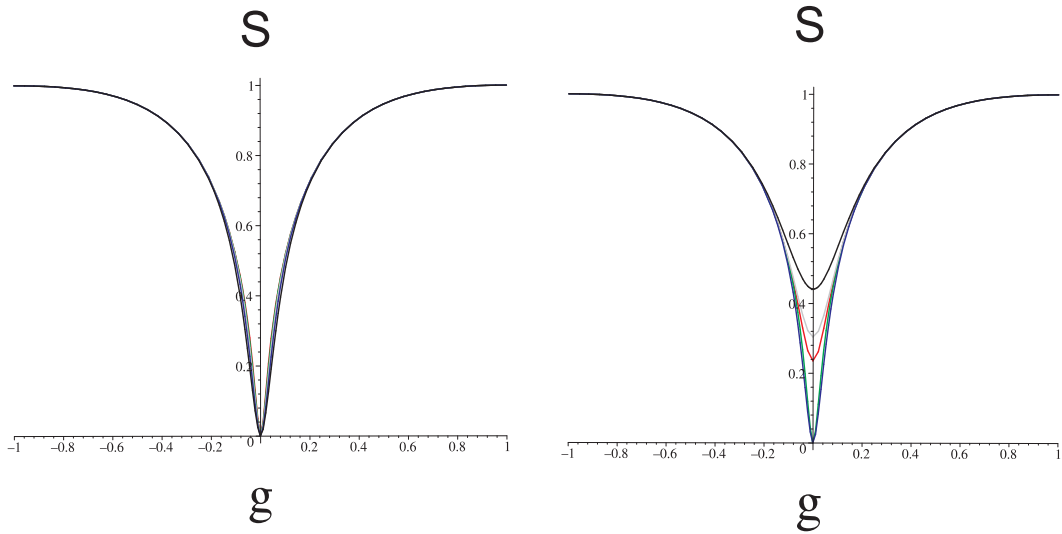}
\caption{ The single site entanglement for models with $\sigma=1$
(left) and $\sigma=-1$ (right), for several values of system
sizes, N=15, 21, 25, 31, and 35. For $\sigma=1$, the entanglement
is almost independent of size, as long as $N>10$ and for
$\sigma=-1$, it is so unless $g$ is very close to the critical
point. }\label{S}
\end{figure}
%%%%%%%%%%%%%%%%%%%%%%%%%%%%%%%%%%%%%%%%%%%%%%%%%%%%%%%%%%%%%%%%%%%%%%%

\subsection{Single site entanglement}
To measure the entanglement of one site with the rest of the
chain we calculate the von Neumann entropy of the density matrix
of one site, which is readily found to be
\begin{equation}
\rho^{(1)}= a(|1\ra\la 1|+|\overline{1}\ra\la
\overline{1}|)+b|0\ra\la 0|
\end{equation}
with
\begin{equation}\label{a}
a=\frac{|g|[(1+2|g|)^{N-1}-(1-2|g|)^{N-1}]}{Z},\h \ b=1-2a
\end{equation}
and
\begin{equation}\label{Zg}
Z=(1+2|g|)^{N}+(1-2|g|)^{N}+2\sigma^N.
\end{equation}
The one-site entropy which measures the entanglement of one site
with the rest of the lattice will then be given by
\begin{equation}\label{Sg}
S=-2a\log(a)-(1-2a)\log(1-2a).
\end{equation}
For $N>>1$ one can verify that $a\simeq g$, therefore for system
sizes, $S(g,N)$ is independent of $N$. Figure (\ref{S}) shows the
behavior of one-site entanglement for different system sizes. It
shows clearly the difference between the models with $\sigma=1$
and $\sigma=-1$. Note that the difference of  $\sigma=-1$ and
$\sigma=1$ models shows up only for systems of odd size.
\subsection{Entanglement of two distant sites}

%%%%%%%%%%%%%%%%%%%%%%%%%%%%%%%%%%%%%%%%%%%%%%%%%%%%%%%%%%%%%%%%%%%%%%%%
\begin{figure}
\centering
\includegraphics[width=6cm,height=6cm,angle=0]{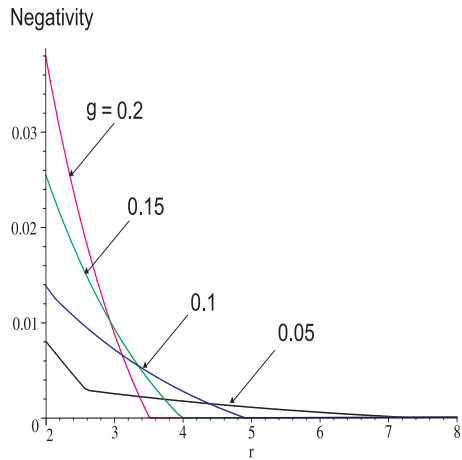}
\caption{ (Color Online) The entanglement of two distant spins as
a function of their distance for a ring of size $N=40$ for several
values of $g$ near the
    critical point.}\label{Neg_r}
\end{figure}
%%%%%%%%%%%%%%%%%%%%%%%%%%%%%%%%%%%%%%%%%%%%%%%%%%%%%%%%%%%%%%%%%%%%%%%

\begin{figure}
\centering
\includegraphics[width=6cm,height=6cm,angle=0]{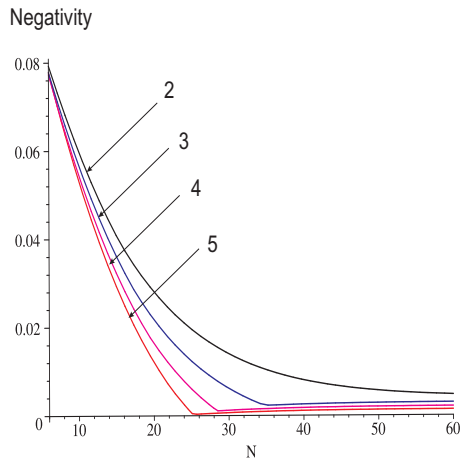}
\caption{(Color Online) The entanglement of two spins with
distances 1, 2 , 3 and 4 (from top to bottom) as a function of
the system size. }\label{Neg_N}
\end{figure}

The entanglement of two sites can be measured by the negativity of
the reduced density matrix of two sites. The general form of
$\rho(1,r)$ is the same as in (\ref{r2Thermo}) except that the
correlation functions now depend on the system size $N$ in a
rather complicated way. We do not write the explicit form of
$\rho(1,r)$ for finite $N$ and only report the behavior of
entanglement in figures (\ref{Neg_r}) and (\ref{Neg_N}). Figure
(\ref{Neg_r}) shows the entanglement of two distant spins as a
function of their distance for a ring of size $N=40$ for several
values of $g$ near the critical point. Several features are
evident. First the entanglement has always a finite range.
Furthermore, the range of entanglement increases as we approach
the critical point, however its value decreases. Near the
critical point, it is well known that correlation lengths
diverge, here we see that entanglement range also diverges
although this is accompanied by lowering of its value. Figure
(\ref{Neg_N}) shows entanglement between adjacent spins as a
function of the size of the system at a fixed value of $g$. The
figure shows that two spins at distance $4$ ($r=5$) can be
entangled for rings of size up to $N=26$. Also for any value of
$g$ and any distance $r$, there is a maximum system size
$N_{max}(r,g)$ above which those two sites can not be entangled
at all. This figure shows that $N_{max}(r,g)$ decreases with $r$.

%%%%%%%%%%%%%%%%%%%%%%%%%%%%%%%%%%%%%%%%%%%%%%%%%%%%%%%%%%%%%%%%%%%%%%%%%
%%%%%%%%%%%%%%%%%%%%%%%%%%%%%%%%%%%%%%%%%%%%%%%%%%%%%%%%%%%%%%%%%%%%%%%

\begin{figure}
\centering
    \includegraphics[width=9cm,height=7cm,angle=0]{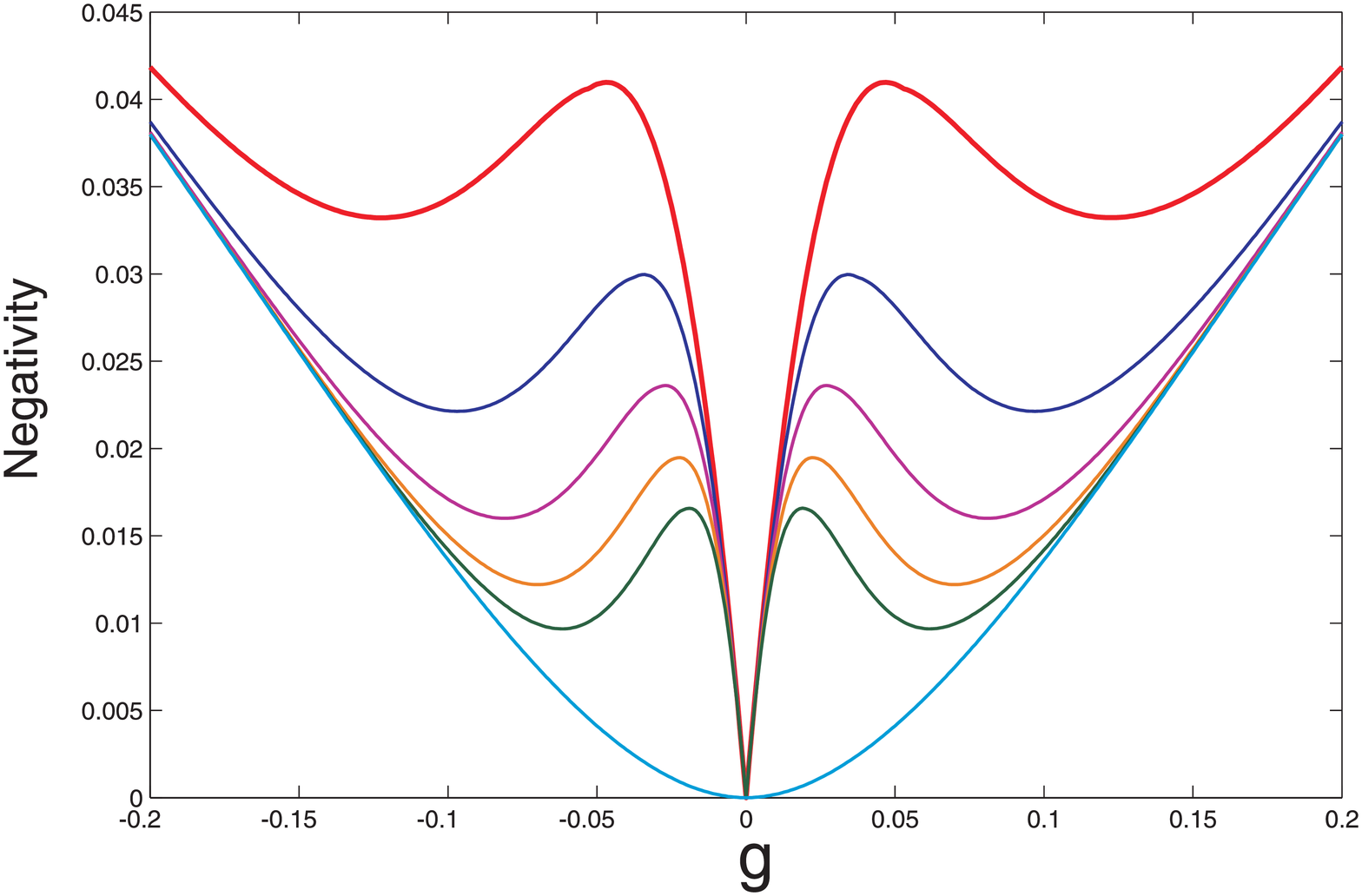}
    \caption{(Color Online) The entanglement of adjacent spins as measured by their
    negativity for different values of system size, from top to bottom equal to
    $N= 15, 20, 25, 30, 35,\infty (\sigma=1)$.}\label{Neg_g}
\end{figure}

\begin{figure}
\centering
    \includegraphics[width=9cm,height=7cm,angle=0]{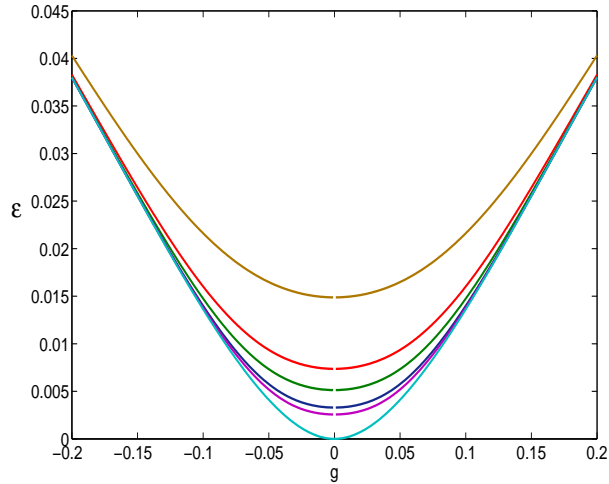}
    \caption{(Color Online) The entanglement of adjacent spins as measured by their
    negativity for different systems with odd-size, from top to bottom equal to
    $N= 15, 21, 25, 31, 35, (\sigma=-1)$.}\label{Neg_gOdd}
\end{figure}

In figure (\ref{Neg_g}) the entanglement of nearest neighbor sites
is displayed for several values of system sizes when $\sigma=1$.
At the critical point, the nearest neighbor spins are not
entangled, however, entanglement appears for any infinitesimal
deviation from this point. On the other hand when $\sigma=-1$
(the class of models which contain the AKLT point), there is
entanglement at $g=0$, only for rings of odd size (figure
\ref{Neg_gOdd}). The behavior for even-sized rings is identical
with the case when $\sigma=1$.\\
%%%%%%%%%%%%%%%%%%%%%%%%%%%%%%%%%%%%%%%%%%%%%%%%%%%%%%%%%%%%%%%%%%%%%%%%
Figure (\ref{Neg_g}) suggests a scaling behavior for entanglement
near the critical point. To investigate this property, we
consider for definiteness, the case $\sigma=1$. For each  value
of $N$, the entanglement attains a maximum at a point $g_m(N)$,
where its value at this point is denoted by $\mathcal{E}_m(N)$.
In the inset of figure (\ref{All2}) we plot $log (g_m)$ and
$log(\mathcal{E}_m)$ versus log (N). We find numerically that
\begin{eqnarray}
  \log (g_m) &=& -1.077 \log (N) - 0.106 \approx - \log N - c,\cr
  \log(\mathcal{E}_m) &=&-1.086 \log N -
  0.214\approx -\log N - d,
\end{eqnarray}
where $c$ and $d$ are two constants, independent of $N$. From the
scaling hypothesis, this means that one can write the negativity
in the vicinity of the critical point as $${\mathcal
E}(g,N)\approx \frac{1}{N}f(Ng),$$ where $f$ is a universal
function. In figure (\ref{All2}) we plot $\log({N\mathcal{E}})$
as a function of  $\log(Ng)$. It shows that all the data
collapse on a single curve for 4 order of magnitudes of $gN$. \\
\subsection{Limiting form of the states}
It is instructive to find the explicit form of the state near the
critical point, i.e. when $|g|\ll 1$. For such an analysis we
should find the dominant amplitudes $\psi_{i_1\cdots i_N}$ in the
linear superposition of all states.
We consider the cases $\sigma=1$ and $\sigma=-1$ separately. \\

\textbf{Case a: $\sigma$=1} Near the critical point $|g|\ll 1$,
the dominant amplitudes are $\psi_{00\cdots 0}$,
$\psi_{k,\overline{l}}$ and $\psi_{\overline{k},l}$, where
$\psi_{k,\overline{l}}$ denotes the amplitude of a state $| k ,
\overline{l}\ra$ in which two spins at sites $k$ and $l$ are
respectively excited to $1$ and $-1$. To first order in $g$, the
state (\ref{state}) becomes
\begin{equation}\label{statea}
    |\Psi\ra\simeq|0,0,\cdots 0\ra
    -g\sum_{k<l=1}^N(|k,\overline{l}\ra+|\overline{k},l\ra).
\end{equation} It is not difficult to find the negativity of this state which is
\begin{equation}
{\mathcal{E}}(\rho(1,r))=2|g|.
\end{equation}
%%%%%%%%%%%%%%%%%%%%%%%%%%%%%%%%%%%%%%%%%%%%%%%%%%%%%%%%%%%%%%%%%%%%%%%%%%
\textbf{Case b: $\sigma$=-1, N=even} For $|g|\ll 1$, the dominant
amplitudes are  $\psi_{00\cdots 0}\propto 1$  and
$$\psi_{k\overline{l}}
\propto
tr(A_0^{k-1}A_1A_0^{l-k-1}A_{\overline{1}}A_0^{N-l})\propto
tr(A_1A_0^{l-k-1}A_{\overline{1}})= 2g(-1)^{l-k},
$$

and $\psi_{\overline{k},l}\propto 2g(-1)^{k-l},$ where we have
used the fact that $A_0^mA_1=A_1$ for any $m$. Thus the state
becomes
\begin{equation}\label{stateb-1}
    |\Psi\ra\simeq|0,0,\cdots 0\ra
    +g\sum_{k<l}(-1)^{l-k}(|k,\overline{l}\ra+|\overline{k},l\ra).
\end{equation}
From the two-site density matrix, one can find the negativity
which is given by $ {\mathcal E}(\rho(1,r))=2|g|$.

\textbf{Case c: $\sigma$=-1, N=odd} In this case we have
$\psi_{00\cdots 0}=0$  and so the state becomes

\begin{equation}\label{stateb1}
    |\Psi\ra=\frac{1}{\sqrt{N(N-1)}}\sum_{k<l}(-1)^{l-k}\left(|k,\overline{l}\ra-|\overline{k},l\ra)\right).
\end{equation}

The entanglement of such a state can be calculated by determining
the reduced density matrix of two sites. After a rather lengthy
but calculation, we find
\begin{eqnarray*}
{\mathcal{E}}(\rho(1,r))&=&\frac{1}{2N(N-1)}|(N-2)(N-3)-1\cr
&-&\sqrt{[(N-2)(N-3)+1]^2+8(N-2)^2}|.
\end{eqnarray*}
\section{Conclusion}
\begin{figure}
  \centering
  \epsfig{file=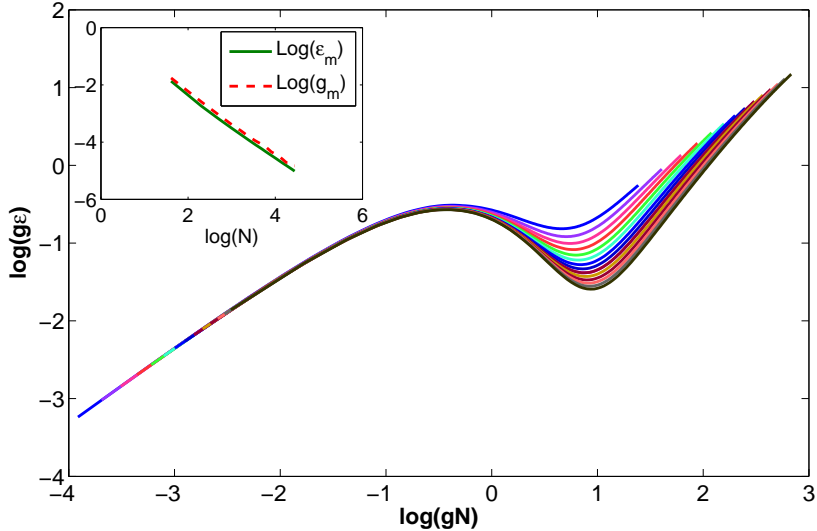,width=12cm}
  \caption{(Color Online) Near the critical point, the entanglement of nearest neighbor sites, as measured by their negativity, shows a scaling behavior. The main
  plot shows $\log(N{\mathcal{E}})$ versus  $log(gN)$, for values of $N=20,25,30 \cdots 80, 85$.
    All the data collapse on one single curve, for a range of 4 orders of magnitude in the value of $gN$. }
  \label{All2}
\end{figure}

In this paper we have studied in detail, the entanglement
properties of a general one-parameter family of matrix product
state of spin one particles defined on a ring. The state has some
plausible symmetries, like rotational symmetry around the $z$,
axis, and the parity symmetry. Such a state is the ground state
of local Hamiltonian defining the nearest neighbor interaction of
spins. The state goes a sharp transition when its continuous
parameter, denoted by $g$, passes through a critical point. In
the thermodynamic limit, this can be ascribed to a quantum phase
transition of the system described by the local Hamiltonian. We
have studied the entanglement properties of the state, near this
transition point, both in the thermodynamic limit and for finite
chains. The study of finite chains is motivated by the possible
role of such systems in quantum computing. We have considered two
measures of entanglement, namely the entanglement of one site
with the rest of the chain, (also studied in \cite{indian} for
infinite rings) and the entanglement of two distant sites with
each other. Our findings can be summarized as follows: 1- When
the parameter $g$ approaches its critical value $g_0=0$, the
range of entanglement (between distant spins) increases
indefinitely at the cost of its magnitude, fig. (\ref{Neg_g_r}),
2- for any non-zero value of $g$, there is a threshold distance
beyond which there is no entanglement between spins, fig.
(\ref{Neg_r}), 3- for any two spins with a fixed distance, there
is a threshold system size, above which entanglement vanishes,
fig. (\ref{Neg_N}), and finally there is a kind of scaling
behavior in entanglement properties of neighboring sites, near the
critical point (Figure \ref{All2}).

{}
\end{document}